\documentclass[11pt]{article}
\usepackage[utf8]{inputenc}
\usepackage{booktabs}
\usepackage{hyperref}
\usepackage{graphicx}
\usepackage{color}
\usepackage{caption}
\usepackage{lineno}
\usepackage{comment}
\usepackage{aas_macros}
\usepackage{amssymb}
\usepackage{verbatim}

\title{Strong Lensing considerations for the LSST observing strategy}

\author{Aprajita Verma, Thomas Collett, Graham P. Smith, \\ 
on behalf of the Strong Lensing Science Collaboration \\ and in collaboration with the DESC Strong Lensing Science Working Group}
\date{November 2018}

\begin{document}

\maketitle

\begin{abstract}
Strong gravitational lensing enables a wide range of science: probing cosmography; testing dark matter models; understanding galaxy evolution; and magnifying the faint, small and distant Universe. However to date exploiting strong lensing as a tool for these numerous cosmological and astrophysical applications has been severely hampered by limited sample sized. LSST will drive studies of strongly lensed galaxies, galaxy groups and galaxy clusters into the statistical age. Time variable lensing events, e.g.\ measuring cosmological time delays from strongly lensed supernovae and quasars, place the strongest constraints on LSST's observing strategy and have been considered in the DESC observing strategy white papers. Here we focus on aspects of ``static'' lens discovery that will be affected by the observing strategy. 
In summary, we advocate (1) ensuring comparable (sub-arcsecond) seeing in the $g$-band as in $r$ \& $i$ to facilitate discovery of gravitational lenses, 
and (2) initially surveying the entire observable extragalactic sky as rapidly as possible to enable early science spanning a broad range of static and transient interests.
\end{abstract}

\section{White Paper Information}
Main corresponding authors:\\
\noindent Aprajita Verma, aprajita.verma@physics.ox.ac.uk\\
Tom Collett, thomas.collett@port.ac.uk\\

\begin{enumerate} 
\item {\bf Science Category:}
\begin{itemize}
\item The Nature of Dark Matter and Understanding Dark Energy
\item Exploring the Changing Sky
\item Galaxy structure and evolution
\end{itemize}

\item {\bf Survey Type Category:} 
Wide-Fast-Deep, Deep Drilling field\footnote{We have included the DDFs in this list as the seeing constraint we are proposing for the WFD equally applies to the DDFs, within which a small number of faint strongly-lensed galaxies are expected to be discovered. It may also be relevant for the DDFs as the $g-$band good seeing advocated for the WFD in this white paper could also impact the seeing distribution split between the DDFs and the WFD.}

\item {\bf Observing Strategy Category:}
    \begin{itemize} 
 
       \item other category:  This paper primarily highlights image quality constraints (seeing $\lesssim 0.8"$) for strong lens discovery.
       It is largely agnostic to the specifics of the short-term observing strategy to build up exposures, although the implications of depth, area, footprint are discussed.
    \end{itemize}  
\end{enumerate}

\clearpage

\section{Scientific Motivation}

Gravitational strong lenses has a broad range of cosmological and astrophysical applications, however, our ability to effectively utilise them is currently hampered by the small number of lenses known. LSST's WFD survey with its combination of wide area, image quality, sensitivity and cadenced observations has the potential to revolutionise strong-lensing science in the 2020s by increasing the number of known strong lenses from order of hundreds to hundreds of thousands (Fig. \ref{fig:lsst-rein}). Such statistically large samples will allow us to realise the huge promise of strong lenses to advance a broad range of fields, including
the detailed mass distributions of lensing galaxies, groups and clusters; new regimes of sensitivity and resolution in high redshift lensed galaxies; testing detailed predictions of CDM on sub-galaxy scales, corresponding to the most likely failure modes of CDM and best constraints on WDM contributions; competitive constraints on cosmological parameters including $H_0$ from lensed quasars and supernovae (e.g. Suyu et al. 2013) and $w$ from double-source plane lenses (Collett \& Auger 2014); and the discovery and study of optical counterparts to strongly-lensed gravitational waves (GWs; Smith et al. 2018a,b see accompanying White Paper by Smith et al.). We expect to discover $\sim 10^5$ galaxy-scale lenses, $\sim 10^4$ strongly lensed quasars and $\sim 500$ lensed type Ia SN (Goldstein \& Nugent 2017) of which 100s will be suitable for time-delay analysis, and 1000s of group- and cluster-scale lenses.

Selecting complete and pure samples of strong lenses is not trivial, mainly owing to the large number of false positives that ‘robots’ (e.g. ArcFinder More et al. 2012 \& RingFinder Gavazzi et al. 2014) and machine learning (ML) codes (e.g. deep convolutional or residual neural networks) currently deliver. Furthermore, none of the lens discovery methods are 100\% complete and are generally optimised to probe different (but overlapping) parameter spaces (in terms of e.g. source brightness, lens redshift, lens mass, and Einstein radius). The SLSC and DESC-SLWG are working towards developing an efficient discovery system for all regimes of strong lensing (galaxy-, group- and cluster-scale), incorporating catalogue and images based selection using robots, sophisticated CNNs and citizen powered visual inspection. However, there are observational considerations (e.g. seeing, depth/area, footprint) that would impact the numbers of lenses discovered in LSST, and therefore impact the resultant science, that are discussed in this white paper. The impact of the observing strategy on cosmography from strongly lensed time delay sources have been summarised in the DESC-WFD and DESC-DDF white papers and are not repeated here.

\begin{center}
\begin{figure}
\begin{minipage}{0.5\columnwidth}
    \centering
    \includegraphics[width=1\columnwidth]{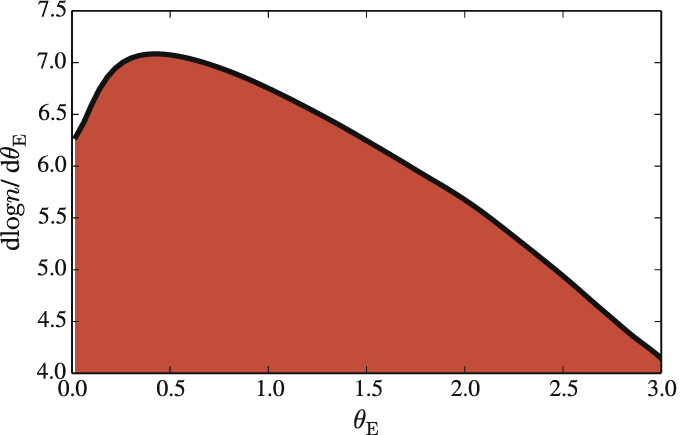}
\end{minipage}
\begin{minipage}{0.5\columnwidth}
    \centering
    \includegraphics[width=1\columnwidth]{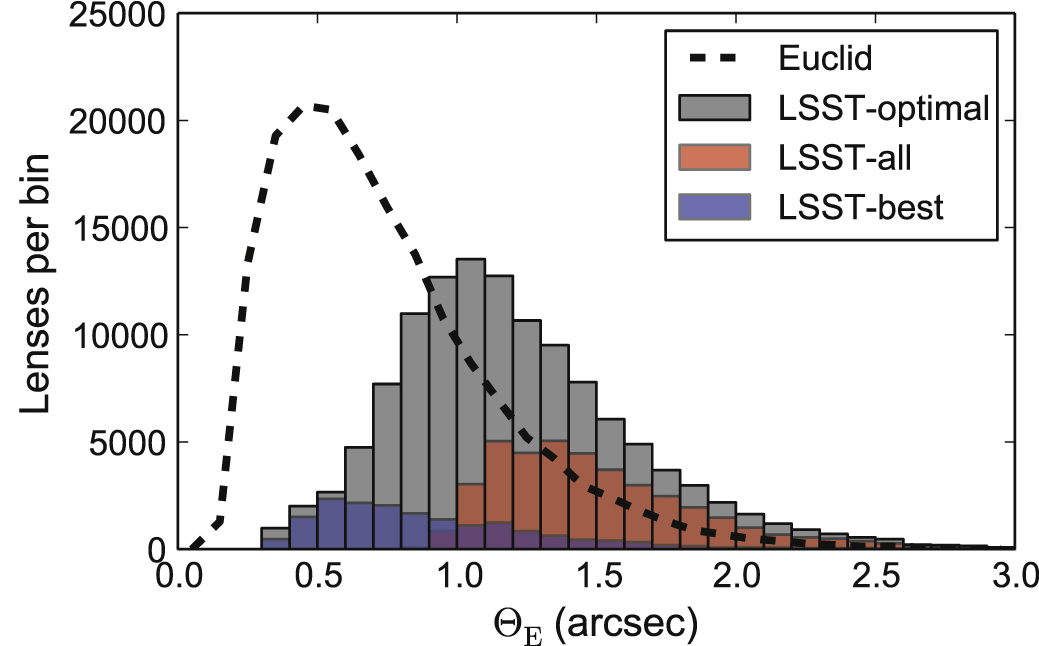}
\end{minipage}
    \captionof{figure}{\footnotesize The expected distribution in Einstein radius ($\Theta_{E}$) of galaxy-galaxy strong lenses theoretically (left) and predicted to be discovered with LSST (right). Grey shows the full LSST lens population, blue is a proxy for the brightest LSST lensed sources and red is the faintest sources (Collett 2015).  The number of lenses we will discover is highly sensitive to the resolution of the images. This is dependent on the seeing (image quality) constraint applied, particularly in the $g$-band, and the final combined image stack that will have a range of actual seeing around the median seeing constraint (see Fig. \ref{fig:lsst-SLsim}). \label{fig:lsst-rein}}
\end{figure}

\begin{figure}
\begin{minipage}{1.\columnwidth}
    \centering
    \includegraphics[width=1\columnwidth]{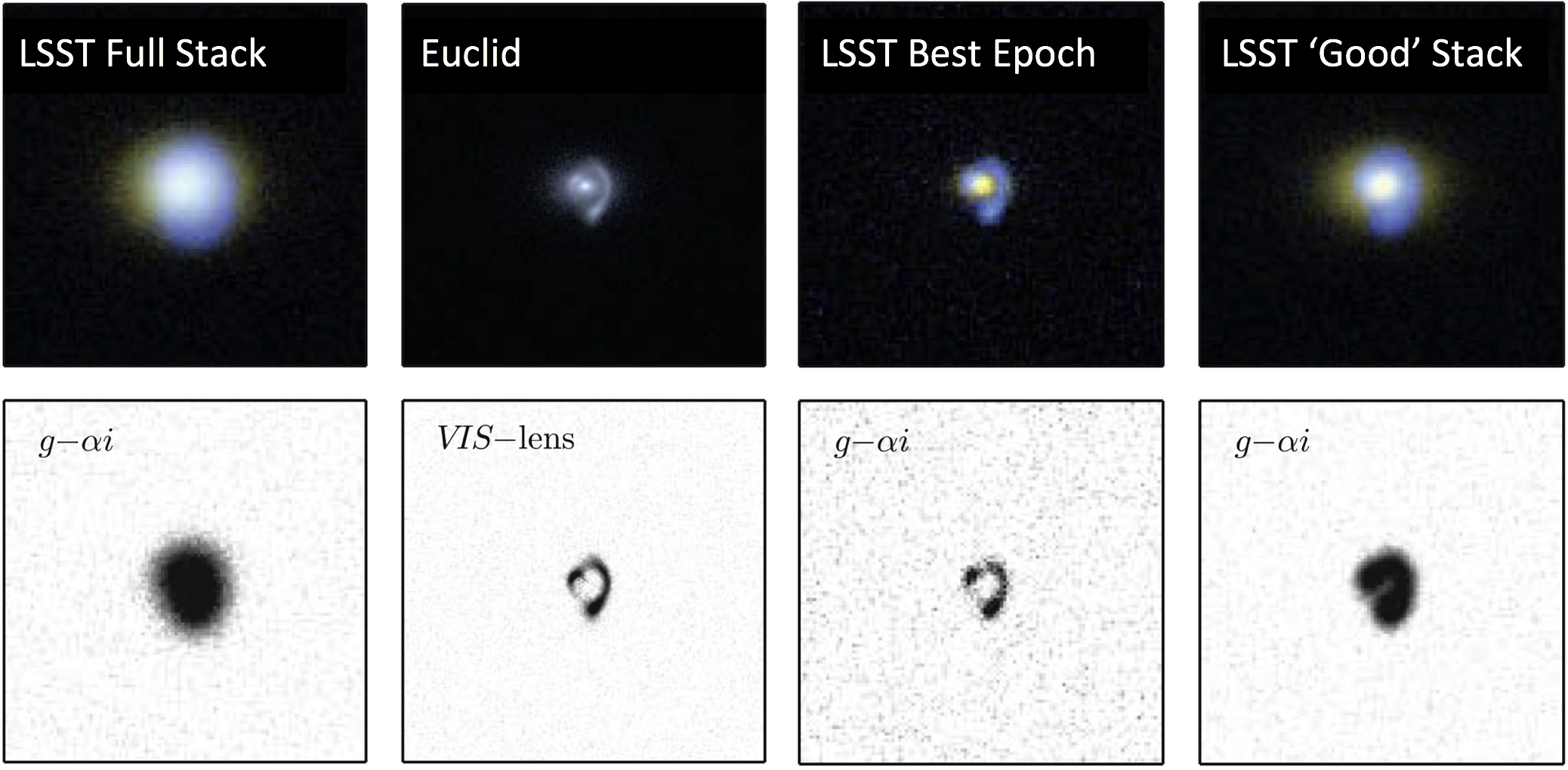}
\end{minipage}
    \captionof{figure}{\footnotesize gri composites for a bright, simulated, galaxy-galaxy, strong lens in typical LSST observing conditions. Left-to-right the images are of the same lens for (a) the full LSST coadded stack, (b) as seen by Euclid, (c) the best single depth LSST exposure and (d) an image stack of only those with 'good' seeing. The bottom row shows the same images but with the light of the lensing galaxy subtracted.  This composite figure shows that it is impossible to identify this system as a lens in the `full stack' which includes exposures taken over a realistic seeing distribution with median seeing $0.8"$. For this bright source, a single `good' seeing exposure is enough to identify this object as being a lens, but most lenses will require multiple good seeing exposures to be coadded, i.e. like the `best' stack (Collett 2015).    \label{fig:lsst-SLsim}}

\end{figure}

\begin{figure}
\begin{minipage}{1\columnwidth}
    \centering
  \includegraphics[width=0.8\columnwidth]{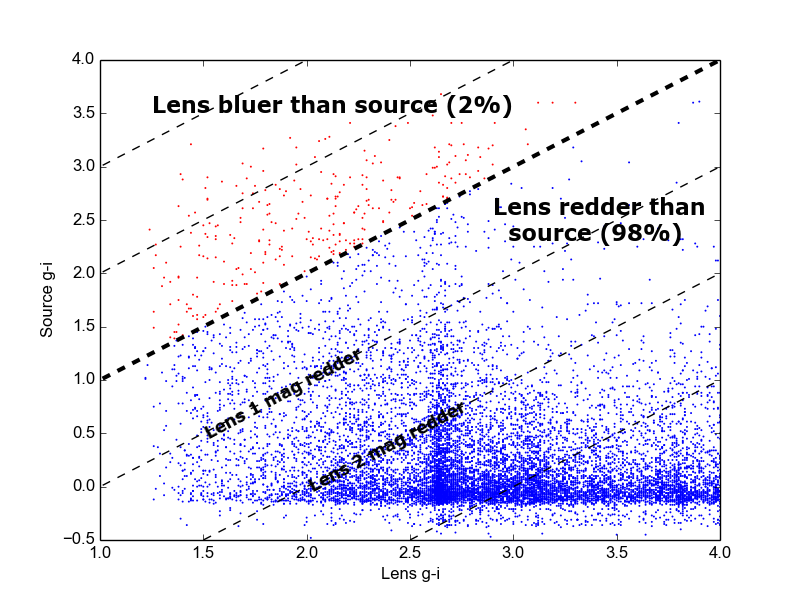}
\end{minipage}
    \captionof{figure}{\footnotesize The majority (98\%) of lensed sources are bluer than the lenses that magnify them, therefore ensuring good blue seeing and sensitivity is paramount to detect the largest number of strongly lensed sources.    \label{fig:lens-colours}}
\end{figure}
\end{center}

\section{Technical Description}\label{sec:tech-intro}

%------------------------------------------------------------------------------------------------------
\subsection{High-level description}\label{sec:lens-discovery}

In general the requirements for strong lens discovery are:
\begin{itemize}
\item \textbf{Wide area with reasonable sensitivity in all bands} increases sample size, the larger the area the more lensed systems will be discovered
\item \textbf{Good (sub-arcsecond) image quality in general, but particularly in the blue (g-band)} to discern faint blue lensed images from the bright red lensing galaxies, better $\Theta_{E}$ sampling, accurate image positions
\item \textbf{Blue sensitivity} detect typically blue SFGs and probe lower down the luminosity function
\end{itemize}

%------------------------------------------------------------------------------------------------------
\subsection{Footprint -- pointings, regions and/or constraints}

Strong lens discovery benefits from wide area with good sensitivity, therefore strategies such as those outlined in the DESC WFD white paper including avoiding areas of high galactic extinction are useful. Overlap with Euclid assists strong lens identification through the availability of high resolution data. The proposed `Peaceful Solutions to the Cadence Wars' white paper also suggests a strategy for a wider area including extension to the north to maximise overlap with e.g. Euclid and DESI. Coverage with wide field spectroscopic facilities such as 4MOST and DESI (see Fig. 3 in the DESC WFD white paper) are useful for redshifts noting the strongly lensed systems will act as filler targets (1-3 per sq. deg.) for the wide field spectrographs. 

Both of these white papers (and others) include the suggestion to move from 2$\times$15s visit pairs to 1$\times$30s, corresponding to a 7\% increase in efficiency. For strong lenses, we would prefer as wide an area of extragalactic sky as possible. An increase of 10\% of the area (with corresponding 10\% decrease in total exposure time) would increase the numbers of strong lenses discovered by $\sim$6\% based on Collett (2015). Having wider area over depth also has the added benefit of increasing the number of brighter lenses discovered which makes spectroscopic follow-up easier.

We request that a full assessment of the costs and benefits of moving to 1$\times$30s exposures is essential. This option results in reduced readnoise due to slower readouts however cosmic ray removal may require detailed simulations and testing. It is not clear what the difference image pair that would populate the alert stream would then be (same night, different filter; same night, same filter; different night same filter). For strongly lensed variable sources, the typical timescales are $>$30s so other transient science cases or bright star saturation concerns in the galactic plane that benefit from t$_{int}<30s$ will have greater influence on this decision.

%------------------------------------------------------------------------------------------------------
\subsection{Image quality}

The predominant population of strongly lensed sources are drawn from the numerous star forming``faint blue galaxy'' population dominating the faint end of the galaxy luminosity function. The predicted distribution of image separations based on simulated galaxy luminosity functions to LSST depths show that the majority of lenses have Einstein radii $\lesssim 0.5"$ (Collett, 2015, Fig. \ref{fig:lsst-rein}) and the intrinsic (unlensed) magnitude of most sources is $i\sim27$. DES chose to optimise observing strategies for weak lensing: prioritising good seeing time in $r$ and $i$ (Diehl et al. 2014). However, making the same choice for LSST would severely hamper the ability to identify strongly lensed galaxy candidates from the LSST data. {\it The very blue nature of lensed sources means that good r-band seeing is no substitute for good g-band seeing.} Simulated OpSim observing strategies that feasibly allow for good median seeing $\sim 0.8"$ in all $gri$-bands with realistic sky brightness distributions (dark time for $g$, and grey/bright time for $r$ and $i$) would ensure that both weak and strong lensing observing requirements can be satisfied. Forecasts using the code of Collett (2015), suggest that relaxing the $g-$band median seeing to $0.9"$ (as in the baseline OpSimv4 and kraken\_2026 sims, as well as \textit{all} the other WP runs) would lead to a significant loss of 17\% of galaxy-galaxy strong lenses and 28\% for $1"$. If the median $g-$band seeing falls to $1.2"$ (similar to the median $g$-band seeing of DES) then 64\% of lenses would be lost.

The current simulated survey observing strategies all have approximately the same median seeing, we request that future simulated runs take account of scientifically driven choices of seeing constraints in scheduling, including the one proposed here. This would presumably also involve the balance of seeing constraints between the WFD, DDFs and other mini-surveys.

%------------------------------------------------------------------------------------------------------
\subsection{Individual image depth and/or sky brightness}
To achieve single visit depths and dependency, we expect $g$-band observations will be carried out under dark time and grey/bright time for $r$ and $i$. Single visit depths (1x30s) over 18,000 square degrees would enable discovery of 100,000 lenses in the most optimistic discovery scenario, or 50,000 using current lens finding methods. These forecasts assume an average of 80x30s visits per pointing in the g-band over the full duration of the survey.

\subsection{Co-added image depth and/or total number of visits}
We prioritise survey area over depth and therefore do not place a strong constraints on the final coadded depth. We refer to the strong lensing case in the DESC WFD paper for total number of visit constraints for strongly lensed SNe and QSOs. 
%------------------------------------------------------------------------------------------------------
\subsection{Number of visits within a night}

For non-variable strong lenses, we are agnostic to the number of visits per night, however we refer to the the DESC-WFD white paper for strongly lensed SNe and QSOs.

%------------------------------------------------------------------------------------------------------
\subsection{Distribution of visits over time}

\subsubsection{A Deep and Early All-sky Reference Imaging Survey}\label{sec:reference}

We propose an ``all-sky reference imaging'' strategy would be a key requirement to enable strong lensing science for a variety of cases. This strategy would ensure that all the available extragalactic sky observed in at least $g$ \& $i$-bands with $\sim 0.8"$ seeing (with an extension to all bands) to single-visit depth at the beginning of the survey operations.  This would be very powerful for a broad range of transient and non-transient science, including discovery of strong gravitational lenses (Section~\ref{sec:lens-discovery}), because single visit $g$ \& $i$-band depth is sufficient for discovery of the brightest strongly lensed sources.  Knowing the location of all the bright lensed sources early on in the WFD survey will allow us to identify lensed supernovae faster, enabling time-critical follow-up to be triggered. Discovery of lensed kilonovae/NS-NS as described in the white paper by Smith et al. will also benefit from this single-visit depth reference imaging strategy.

For non-variable strong lenses, we are agnostic to how the remaining depth is collected, however uniform full sky/footprint coverage tied to the annual data releases would be preferred to maximise the discovery of lenses (in line with the strong lensing recommendation in the DESC WFD).

%------------------------------------------------------------------------------------------------------
\subsection{Filter choice}

Coverage in all bands would be preferable to enable accurate photo-z estimates and spectral template fitting to characterize both the lens and source, although our selection will largely be based on $gri$ imaging for which we request good image quality and sensitivity.

\subsection{Exposure constraints}
No constraints.

%------------------------------------------------------------------------------------------------------
\subsection{Other constraints}
No other constraints.

%------------------------------------------------------------------------------------------------------
\subsection{Estimated time requirement}\label{sec:time}
This science will be carried out in the WFD survey. Our default estimates are based on, 80x30s exposures in $g$ across 18,000 square degrees with a median seeing of $0.81"$. will yield 100,000 lenses. Comparable depth and image quality in at least $i$ is also required.

%------------------------------------------------------------------------------------------------------

%\vspace{.3in}
\begin{table}[ht]
    \centering
    \begin{tabular}{|l|l|}
        \toprule
        Properties & Importance \\
        \midrule
        Image quality  &  1 (g especially) \\
        Sky brightness & 1 \\
        Individual image depth & 2  \\
        Co-added image depth & 2  \\
        Number of exposures in a visit   & 3  \\
        Number of visits (in a night)  & 3  \\ 
        Total number of visits &  3 \\
        Time between visits (in a night) & 3\\
        Time between visits (between nights)  & 3  \\
        Long-term gaps between visits & 3\\
        \bottomrule
    \end{tabular}
    \caption{{\bf Constraint Rankings:} Summary of the relative importance of various survey strategy constraints. Please rank the importance of each of these considerations, from 1=very important, 2=somewhat important, 3=not important. If a given constraint depends on other parameters in the table, but these other parameters are not important in themselves, please only mark the final constraint as important. For example, individual image depth depends on image quality, sky brightness, and number of exposures in a visit; if your science depends on the individual image depth but not directly on the other parameters, individual image depth would be `1' and the other parameters could be marked as `3', giving us the most flexibility when determining the composition of a visit, for example.}
        \label{tab:obs_constraints}
\end{table}

%------------------------------------------------------------------------------------------------------

\subsection{Technical trades}

    The number of strong lenses discovered is improved through increased area over depth. For example, based on Collett (2015), 10\% more extragalactic area, (and a corresponding 10\% lower exposure time across the field) increases the number of strong lenses discovered by 6\%. While we note this, the strong lensing case does not provide a strong constraint on the depth vs. area debate. We will discover lenses in all scenarios ($\pm$ 10\%) being proposed for the WFD, we do however express a preference for as large an area as possible to be covered. The non-variable strong lens case does not place any strong constraint on the trades between exposure time and number of visits, or uniformity and coadded depth. The tolerance on the image quality (seeing) of images used in the coadd would be of more relevance to the strong lensing discovery case than optimised real-time exposure time.   Given the stronger constraints on trades comes form variable strong lenses, we refer to the DESC WFD white paper.

%------------------------------------------------------------------------------------------------------

\section{Performance Evaluation}

Our science goals will be achieved in full if median $g$-band seeing is $0.8"$. Nights with g-band seeing $ << 0.8"$ are particularly important. Ideally we would request the majority of the good seeing time be allocated to $g$. Since such a request would be in great tension with other science, we simply request that $g$ band seeing is not compromised by prioritizing seeing in other bands. 
If median g-band seeing drops to $0.9"$ we will lose 17\% of lenses. If median $g$-band seeing is $1"$ or worse our science goals would be greatly diminished.
 
Even in $g$, the median LSST lensed source is still 2 magnitudes fainter than the lensing galaxy. In the r-band the median source is 4 magnitudes fainter than the lens. {\it We cannot achieve our goals with good r-band seeing in lieu of good g-band seeing.}

%------------------------------------------------------------------------------------------------------

\section{Special Data Processing}

    As shown in Fig. \ref{fig:lsst-SLsim}, using the LSST full stack, the optimal or `good' seeing stack and/or the best seeing single epoch image will heavily influence our ability to (a) discover lenses (b) extract accurate LSST photometry for the lens and the source. We therefore anticipate needing additional processing to produce `good' seeing stacks, as these will not be produced as part of the baseline data products. We will pre-select luminous red galaxies using the classifications of the LSST DM deblender in level 2 images around which we will produce seeing optimised stacks and also select the `best' single visit. We expect such a selection will amount to 10$^3$ times as many cutout images as expected lenses by the end of the 10 year survey (i.e. of order 10$^{9-10}$ single-depth image cutouts will be needed in at least $g$, $r$ \& $i$ to produce such seeing optimised stacks). We do not yet have plans on how to achieve this, but will not need all data simultaneously to produce the stacks which can be done in batches and the final stacks only stored. We will use simulations to assess and develop such software ahead of commissioning data, and also investigate potential level 3 techniques with the DM deblender(s) and lens subtraction (e.g. BlueRings, Collet et al. in prep) to develop strategies to efficiently find lenses in the level 2 stacks and generate the required data products and image cutouts.

%------------------------------------------------------------------------------------------------------

\section{References}

Collett \& Auger, 2014, MNRAS, 443, 969\\
Collett 2015, ApJ, 811, 20\\
Diehl et al., 2014, Proc. SPIE, 9149, 91490V\\
Gavazzi et al., 2014, ApJ, 785, 144\\
Goldstein \& Nugent, 2017, ApJ, 834, L5\\
More et al., 2012, ApJ, 749, 38\\
Smith G. P., Jauzac M., et al., 2018a, MNRAS, 475, 3823\\
Smith G. P., Bianconi M., et al.\ 2018b, arXiv:1805.07370\\
Suyu et al., 2013, ApJ, 766, 70\\

\end{document}